\documentclass[9pt,twocolumn,twoside]{pnas-new}

\templatetype{pnasresearcharticle} 

\usepackage{bm}
\usepackage{stmaryrd}

\title{Global method for gender profile estimation from distribution of first names}

\author[a,b]{Manolis Antonoyiannakis}
\author[c,d]{Hugues Chat\'e}
\author[a]{Serena Dalena} 
\author[a]{Jessica Thomas}
\author[a,1]{Alessandro S. Villar}

\affil[a]{American Physical Society, 1 Physics Ellipse, College Park, 20740 Maryland, USA}
\affil[b]{Department of Applied Physics \& Applied Mathematics, Columbia University, New York, USA}
\affil[c]{Service de Physique de l'Etat Condens\'e, CEA, CNRS, Universit\'e Paris-Saclay, CEA-Saclay, 91191 Gif-sur-Yvette, France}
\affil[d]{Computational Science Research Center, Beijing 100193, China}

\leadauthor{Antonoyiannakis} 

\significancestatement{
As social issues related to gender bias attract closer scrutiny, 
accurate tools to determine the gender profile of large groups become essential. 
When explicit data is unavailable, gender is often inferred from names.
Current methods follow a strategy whereby individuals of the group, one by one, are assigned a gender label or probability based on gender-name correlations observed in the population at large. 
We show that this strategy is logically inconsistent and has practical shortcomings, the most notable of which is the systematic underestimation of gender bias. 
We introduce a global inference strategy that estimates gender composition according to the 
context of the full list of names. 
The tool suffers from no intrinsic methodological effects, is robust against errors, easily implemented, and computationally light.
}

\authorcontributions{ASV devised the global method; MA tested the gender inference methods; HC, SD, and JT devised ways to test the global method; HC and ASV wrote the paper.}
\authordeclaration{The authors declare no conflict of interest.}
\correspondingauthor{\textsuperscript{1}To whom correspondence should be addressed. E-mail: villar@aps.org}

\keywords{gender profile inference $|$ global gender estimation method $|$ conditional probabilities estimation $|$ gGEM}

\begin{abstract}
Current approaches to infer the gender profile of a group exploit empirical correlations observed in the 
population at large to guess, one by one, the likely gender of every name in the group. 
We show that such `individual-based gender estimation methods' (iGEMs) are logically inconsistent due 
to implicit reliance on a fair sampling assumption. 
Moreover, their gender estimates are intrinsically biased, systematically overestimating the participation of the minority gender. 
We introduce an inference strategy based on a global and self-consistent analysis of the target list of names that, 
by relaxing the fair sampling assumption and taking into account contextual information, 
aims at optimal gender classification. 
Our `global gender estimation method' (gGEM) relies on a leaky pipeline model of social dynamic and 
is inherently devoid of the logical inconsistencies and systematic errors of iGEMs.
We employ artificially-generated populations of varying gender compositions to benchmark gGEM against iGEMs.
gGEM achieves high accuracy and robustness against misclassification of individual names.
The method outperforms iGEM approaches, particularly for populations showing high degree of gender bias or large numbers of unisex names. 
In fact, gGEM is able to produce accurate gender profile estimates even when relying on weak proxies, such as first-name initials. 
The leaky pipeline model at the heart of gGEM provides a quantitative and intuitive dynamical perspective on the social processes causing gender imbalance.
\end{abstract}

\dates{This manuscript was compiled on \today}
\doi{\url{www.pnas.org/cgi/doi/10.1073/pnas.XXXXXXXXXX}}

\begin{document}

\maketitle
\thispagestyle{firststyle}
\ifthenelse{\boolean{shortarticle}}{\ifthenelse{\boolean{singlecolumn}}{\abscontentformatted}{\abscontent}}{}

\dropcap{H}ow to determine the gender makeup of a population when gender information is not available is a long-standing problem that has become more important with increased focus on understanding gender bias issues.
(For recent examples in STEM publishing only, see e.g. \cite{helmer2017research,king2017men,murray2019author,huang2020historical,dworkin2020extent,chatterjee2021gender,squazzoni2021peer,teich2022citation,ross2022women}.)
First names, a form of public identification considered of low sensitivity when it comes to privacy, are well correlated with two gender identities (women and men) in the population at large, 
making them a practical proxy for estimating the relative participation of these two genders in a group\footnote{Other gender identities are not known to show clear correlations with names and thus cannot have their relative participation estimated from gender-name inference strategies in general.}. 

A significant roadblock for gender-name inference strategies stems from the fact that first names are often not perfectly allocated to one gender, i.e. may be unisex.
For example, about 18\% of all individuals ever registered in the U.S. Social Security Administration baby name public database~\cite{dataset_US} bear first names with probability larger than 1\% of not belonging to the majority gender. 
Although the imprecision may appear small and thus inconsequential, 
the compounded effect of small errors in the whole population, as we show in this paper, becomes significant when targeting 
highly gender-skewed groups.

Existing methods of gender inference 
assign a gender label (e.g. male, female, undefined) or a gender probability to every name of interest 
considered in isolation.
Most only use first names and rely on publicly or commercially available statistics 
extracted from the population at large
\cite{lariviere2013bibliometrics,wais2016gender,karimi2016inferring,santamaria2018comparison,fortin2021digital},
but some employ sophisticated machine learning or multi-factor approaches considering other pieces of 
information~\cite{das2021context,hu2021what,smith2013search,ethnea2016torvik,muller2017improving,vanbuskirk2022open}. 
Because in these methods the gender profile of a group is derived essentially from counting the gender labels assigned to each person given the name, 
we refer to them as `individual-based gender estimation methods' (iGEMs).

iGEM strategies rely on an implicit assumption of \textit{fair sampling}, i.e., that the group under scrutiny is a typical 
sample of the population at large on which gender labels are based. 
However, this assumption generally breaks down for groups of interest because of the gender imbalance caused by gender bias. 
We show that iGEMs incur systematic errors in gender profile estimates, 
particularly for populations that are highly gender-unbalanced and/or 
including significant amounts of unisex names. 

Consider, as an illustration of principle, a list of names sampled ---unbeknownst to us--- from the members of a gentlemen's club. 
Suppose that the name ``Carol'', which belongs to a woman with about 99\% probability in a typical sample of the population~\cite{dataset_US}, 
appears on the list. 
In this case, classifying individuals based on statistics borrowed from the population at large, and not on the \textit{best} contextual information,
would mistakenly indicate the presence of at least one woman in this club. Unisex names will make the problem even worse. 

Having realized that associating a gender label to names in isolation entails an intrinsic methodological limitation, 
we introduce a ‘global Gender Estimation Method’ (gGEM) that relaxes the assumption of fair sampling. 
This is done by exploiting information available not only from the population at large, but from the context provided by the 
list of names of the population of interest, and then estimating gender-name correlations within the population itself\footnote{See~\cite{ggemapp} for a web app implementation of gGEM.}.
Compounding the partial information provided by each name, the method produces an accurate description of the whole ensemble, 
free of intrinsic methodological systematic errors. We therefore demonstrate how the distribution of names can provide contextual 
information that allows better gender guesses.

\section*{Individual-Based Gender Estimation Methods (iGEMs)} 

Gender profile inference aims to estimate $N_g$, the number of people with gender $g\in\{f,m\}$\footnote{Where $f$ and $m$ stand for 'female' and 'male'. 
Gender inference by names is limited to gender classes strongly correlated with sex; We choose to highlight this limitation by using 'female' and 'male' as labels, 
even though the stated goal of our inference method is to provide a tool to characterize the relative presence of women and men. }
in a \textit{target population} $\mathbf{T}$ for which the only information available is the number of individuals $N(s)$ bearing the first name~$s$.

The most common strategy found in the literature 
involves assigning a gender probability to each name $s$ in $\mathbf{T}$ using the proportions observed in a 
reference population $\mathbf{R}$. 
Each \textit{conditional probability} $p_R(g|s)$ that a person 
in $\mathbf{R}$ with first name $s$ has gender $g$ must be retrieved 
from an independent source, which is typically a dedicated commercial service or a publicly available database such as a national census, 
and may involve sophisticated data processing~\cite{das2021context,hu2021what,smith2013search,ethnea2016torvik,muller2017improving,vanbuskirk2022open}. 

Such iGEM strategies implicitly assume that conditional probabilities $\{p_R(g|s)\}$ are appropriate to describe the target population $\mathbf{T}$. 
A simple iGEM would consist in choosing, as a possible estimate
\begin{equation}
N^{\mathrm{(0)}}_g = \sum_{s} p_R(g|s)N(s), 
\label{eq:method_0}
\end{equation}
where the superscript ``(0)'' labels this particular implementation of iGEM. 

This form of estimation is not logically consistent. 
As a sample of the human population at large, $\mathbf{R}$ is close to gender balance, a property reflected on the set of conditional probabilities $\{p_R(g|s)\}$. 
The gender profile of $\mathbf{T}$, on the other hand, is assumed unknown (hence the need to estimate it in the first place).
A principled estimate of gender composition of $\mathbf{T}$ would require knowledge of the conditional probabilities $\{p_T(g|s)\}$ instead, 
reflective of the (unknown) gender-name correlations particular to this population\footnote{The number of people with gender $g$ in $\mathbf{T}$ is, by definition, $N_g = \sum_{s} p_T(g|s)N(s)$. }. 

Eq.~(\ref{eq:method_0}) must thus be understood as an approximation, namely that $p_R(g|s)\approx p_T(g|s), \forall s$. 
The approximation is reasonable in two scenarios: 
\begin{description}
\item[(i)]{If $p_R(f|s)\approx1$ or $p_R(m|s)\approx1$, since names with strong gender association remain the most unaffected by differences in gender composition between $\mathbf{R}$ and $\mathbf{T}$, or}
\item[(ii)]{If $\mathbf{T}$ is close to gender-balance, in which case it can be considered a fair sample of $\mathbf{R}$ for estimation purposes.}
\end{description}
Outside of these domains, Method (0) is not expected to produce reliable estimates. We later investigate its limitations quantitatively. 

A possible improvement of Method~(0) involves introducing a cutoff probability $p_c$, usually in the range of 70\% to 90\%, to pre-select names fulfilling scenario~(i) in the form $p_R(f|s)\geq p_c$ or $p_R(m|s)\geq p_c$. 
Names which do not fulfill this condition are discarded and do not enter the gender estimate. 
This ``Method~(1)'' would restrict the sum in \eqref{eq:method_0} to a subset of names with `well-defined' gender association:
\begin{equation}
N^{(1)}_g = \sum_{\{s \,|\, p_R> p_c\}} p_R(g|s)N(s)
\label{eq:method_i}
\end{equation}
(see, e.g.~\cite{king2017men,thomas2019gender,dworkin2020extent,mattauch2020bibliometric,squazzoni2021peer,chatterjee2021gender,dew2021gendered,teich2022citation} for recent examples).
Note that this equation reduces to \eqref{eq:method_0} if $p_c\leq50\%$.

This restriction, while producing more accurate estimates, has the disadvantage of still being logically inconsistent if $p_c$ is not very close to 100\%. 
In practical terms, the restriction disregards people bearing unisex names from the estimate, thus decreasing the quality of sampling and increasing statistical uncertainty. 
This effect can be especially relevant for groups from East Asia, due to higher prevalence of unisex names~\cite{huang2020historical}. 
It is also not clear from the literature what magnitude of errors should be expected as a consequence of the approximation. 
We investigate this issue numerically to show that Method~(1), i.e., Eq.~(\ref{eq:method_i}), mitigates but does not exclude systematic errors, particularly for populations with high degree of gender imbalance. 

Most iGEM variants replace the conditional probabilities in Eq.~(\ref{eq:method_0}) by a maximum likelihood rule. 
If $p_R(g|s)>p_c$, then all $N(s)$ individuals are classified as belonging to gender $g$ [see \cite{helmer2017research, king2017men, murray2019author, dworkin2020extent, chatterjee2021gender, teich2022citation, lariviere2013bibliometrics, fortin2021digital, thomas2019gender, dew2021gendered} for examples]. 
Once more, names for which the condition is not satisfied are disregarded. This variation, which we refer to as Method~(2), produces the estimate: 
\begin{equation}
N^{(2)}_g = \sum_{s} \Theta [p_R(g|s)-p_c] N(s),
\label{eq:method_ii}
\end{equation}
where $\Theta [x] =1 $ if $x>1$ and $\Theta [x]=0$ if $x\leq0$. 
Method (2) entails similar shortcomings to those of Method~(1) with quantitatively smaller errors.

\section*{Global Gender Estimation Method (gGEM)}

We introduce an inference strategy that bypasses the need for precise gender classification of individual names in isolation 
to focus instead on the population as a whole from the start. 
We aim to find the `best' gender classification to every name in the \textit{context} of the observed list, according to a certain procedure. 

At this stage, it is helpful to define global parameterizations of the gender composition of $\mathbf{T}$. 
We consider three equivalent parametrizations. 
The global ratio $\alpha$  of females to males in $\mathbf{T}$, $\alpha\in[0,\infty[$, is defined as 
\begin{equation}
\alpha := \frac{N_f}{N_m}\;.
\label{eq:def_alpha}
\end{equation}
The equivalent quantity for the pristine population $\mathbf{R}^*$ is $\alpha^* := N_f^*/N_m^*$. 
Since $\mathbf{R}^*$ is assumed a fair sample of the population at large $\mathbf{R}$, which is gender-balanced (i.e., $N_f^*\approx N_m^*$), 
we assume $\alpha^*=1$ to simplify derivations in what follows. 

The fraction of female individuals $\beta$ (with $0\leq\beta\leq1$), perhaps the most intuitive global parameter, is defined as
\begin{equation}
\beta := \frac{N_f}{N_f+N_m} = \frac{\alpha}{1+\alpha}\;.
\label{eq:def_beta}
\end{equation}
Finally, we define the gender imbalance $\gamma$ (with $-1\leq\gamma\leq1$), representing the departure from gender-parity, as
\begin{equation}
\gamma := \frac{N_f-N_m}{N_f+N_m} = \frac{\alpha-1}{\alpha+1}.
\label{eq:def_gamma}
\end{equation}
For a gender-balanced population, $\alpha=1$, $\beta=\frac{1}{2}=50\%$, and $\gamma=0$. 
For a male(female)-only population, $\alpha=0$, $\beta=0$, and $\gamma=-1$ ($\alpha\rightarrow\infty$, $\beta =1=100\%$, and $\gamma=1$). 

The goal of gender estimation methods is to provide an accurate and robust guess for these parameters.

\subsection*{Leaky pipeline model of social dynamic}

How gender bias emerges from social processes is often described as a `leaky pipeline':
people of a particular gender experience disadvantages while navigating through a series of selective steps.
One of the aggregated effects of such social dynamic is gender imbalance. 

gGEM implements in mathematical terms the very idea that the target population $\mathbf{T}$ may be thought of as originating 
from a typical subset of the population at large $\mathbf{R}$, represented as $\mathbf{R}^*$, through an aggregated gender-dependent social process.
A leaky pipeline social dynamic will affect the frequency of a given name $s$ in a way that is sensitive to its gender association, according to the expressions
\begin{align}
\label{eq:def_transf_biased_pop_1}
N_f(s) & \; = c_f N^*_f(s), \\
N_m(s) & \; = c_m N^*_m(s), 
\label{eq:def_transf_biased_pop_2}
\end{align}
or simply $N_g(s)=c_g N^*_g(s)$. 
The constants $c_g\leq1$ ---which do not depend explicitly on names--- represent the relative `loss' of people of gender $g$. 
They characterize how the leaky pipeline transforms a hypothetical `pristine' group of people $\mathbf{R}^*$ comprised of 
$\{N^*_f(s)\}$ females and $\{N^*_m(s)\}$ males (babies belonging to the population at large, if one wishes), into the observed 
population $\mathbf{T}$ at a different point in time, comprised of $\{N_f(s)\}$ 
females and $\{N_m(s)\}$ males.

\subsection*{Transformation of conditional probabilities}

As a consequence of the leaky pipeline, gender-name conditional probabilities in $\mathbf{T}$ depend both on the 
gender mix of the name in $\mathbf{R}$ and on the \emph{gender dependence of the social dynamic} through the coefficients $c_g$. 
Using Eqs.~(\ref{eq:def_transf_biased_pop_1})--(\ref{eq:def_transf_biased_pop_2}), we substitute $N_g(s)$ and $N(s)$ in the conditional-probabilities identity, $p_T(g|s) = N_g(s)/ N(s)$, to obtain the conditional probabilities in $\mathbf{T}$ as 
\begin{align}
\label{eq:relation_condprob_1}
p_T(f|s) & = \frac{\eta \, p_R(f|s)}{\eta \, p_R(f|s) + p_R(m|s)}, \\
p_T(m|s) & = \frac{ p_R(m|s)}{\eta \, p_R(f|s) + p_R(m|s)},
\label{eq:relation_condprob_2}
\end{align}
with $\eta = c_f/c_m$, 
where we used the equivalent identities $N^*_g(s) = p_R(g|s) N^*(s)$ for $\mathbf{R}^*$. 
The parameter $\eta$ represents the ratio of female to male loss through the pipeline. 
Hence $\eta<1$ (or $\eta>1$) represents a leaky pipeline for females (or males), whereas $\eta\approx1$ entails a fair or equitable pipeline.

\begin{figure}[tb]
\center{\includegraphics[width=0.8\columnwidth]{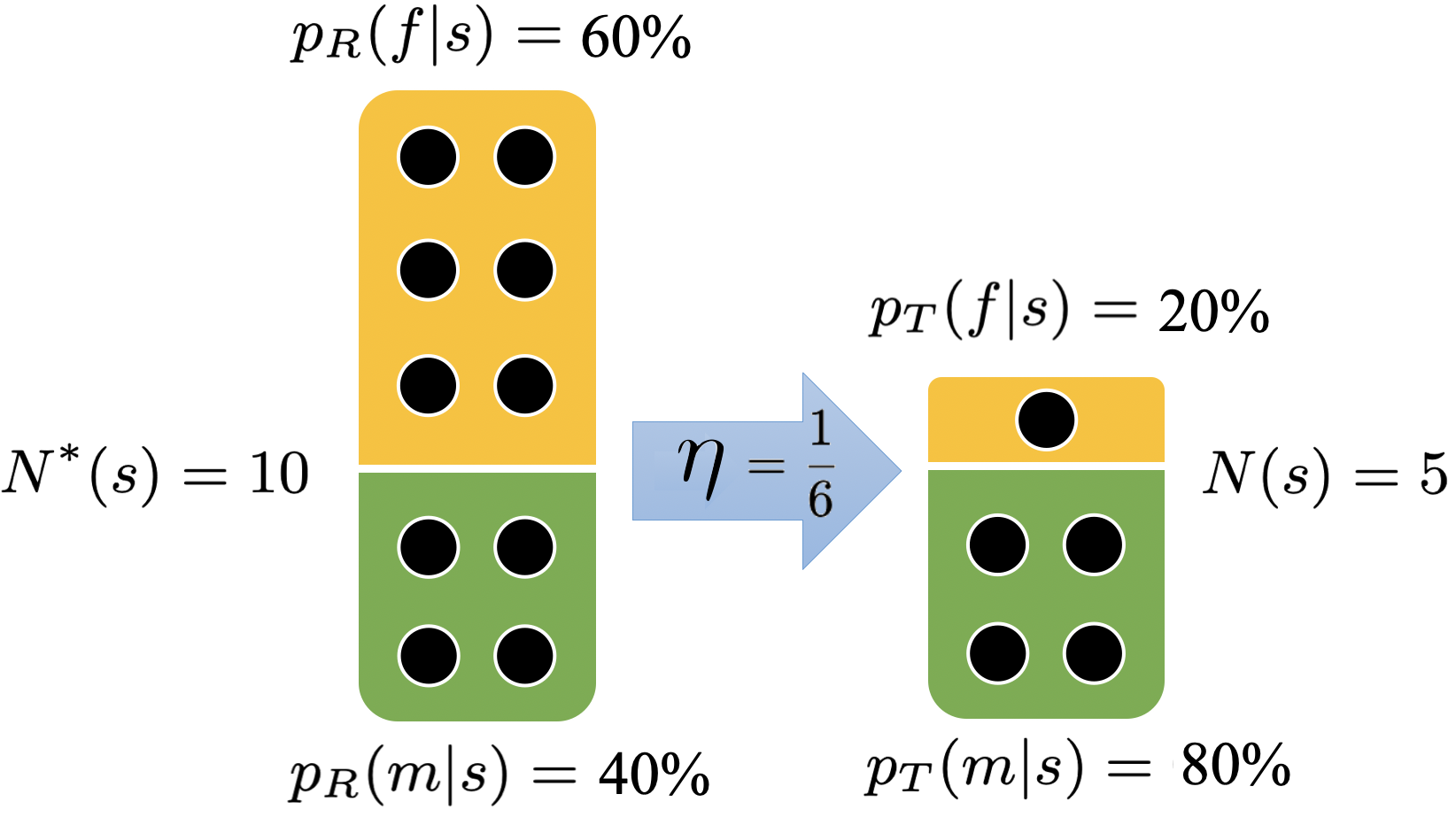}}
\caption{\label{fig:illustration_cond_prob} Illustration of the net effect of a gender-dependent social process on the conditional probabilities that guide the gender attribution of names in a population. The areas of the rectangles represent the relative fractions of 
female (yellow, top) and male (green, bottom) individuals named $s$ in population $\mathbf{R}$ (left) and $\mathbf{T}$ (right). A `leaky pipeline' social dynamic removes in this example a relative 
fraction $\eta=1/6$ of all females. The conditional probabilities favoring `female' as gender classification of 10 individuals (black circles) in $\mathbf{R}$ (since $p_R(f|s)=0.6>0.4 =p_R(m|s)$ there), are actually very likely to belong to a male individual in $\mathbf{T}$, since then $p_T(m|s)=0.8>0.2=p_T(f|s)$ as a result of the social dynamic.}
\end{figure}

The transformation of Eqs.~(\ref{eq:relation_condprob_1})--(\ref{eq:relation_condprob_2}) has a simple interpretation, illustrated in Fig.~\ref{fig:illustration_cond_prob}. 
Considering a name $s$ in $\mathbf{R}$ with associated probabilities $p_R(g|s)$, if the leaky pipeline produces a relative change $\eta$ in the female-to-male proportion, the probability of finding a female individual under $s$ is simply given by the new proportion of females $\eta p_R(f|s)$ normalized by the corresponding relative change in the total number of individuals bearing the name, $\eta p_R(f|s)+p_R(m|s)$. 

Eqs.~(\ref{eq:relation_condprob_1})--(\ref{eq:relation_condprob_2}) thus yield the conditional probabilities corrected to reflect gender-imbalance in $\mathbf{T}$ as presumed by $\eta$. 
These conditional probabilities represent the best gender-name guess \textit{updated by the knowledge} that gender participation changes by a relative amount $\eta$ with respect to the pristine population $\mathbf{R}^*$. 

\begin{figure}[!tb]
\centering
\center{\includegraphics[width=0.99\columnwidth]{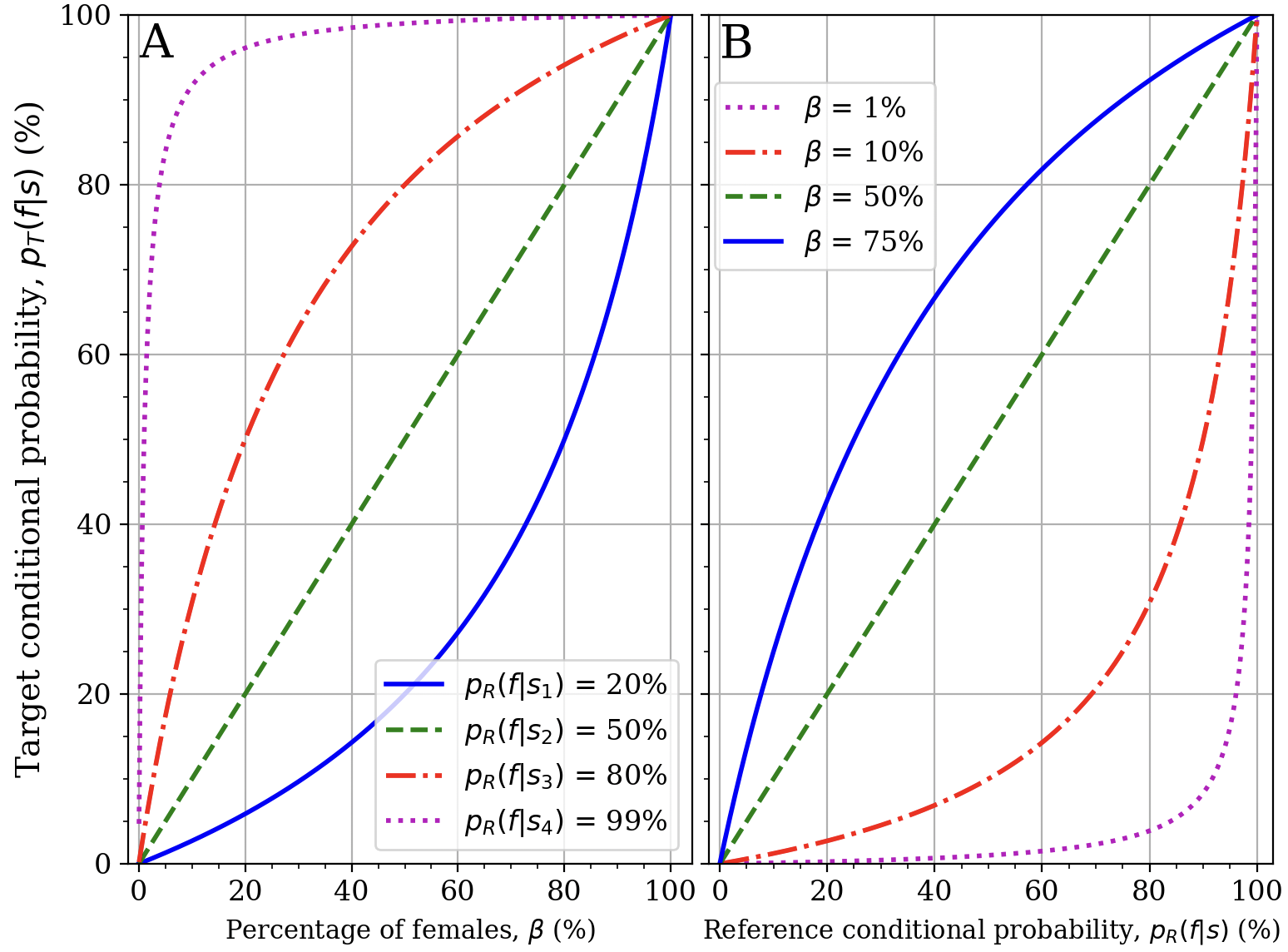}}
\caption{\label{fig:cond_prob_functions} Transformation of female-name probabilities from $\mathbf{R}$ to $\mathbf{T}$ induced by the social dynamic that produces a gender 
composition $\beta=N_f/(N_f+N_m)$ in $\mathbf{T}$, according to Eq.~(\ref{eq:relation_condprob_1}). 
\emph{(A)} Female-name probabilities in $\mathbf{T}$ as functions of $\beta$ for four names 
with very different nominal probabilities in $\mathbf{R}$. 
\emph{(B)} Conversion between female probabilities in $\mathbf{R}$ and $\mathbf{T}$ for 
various gender compositions $\beta$ of the target group.}
\end{figure}

Figure~\ref{fig:cond_prob_functions} depicts a few examples of how conditional probabilities in $\mathbf{R}$ 
are transformed to describe a population $\mathbf{T}$ with female fraction $\beta$. 
From panel \emph{(A)}, conditional probabilities can be seen to remain unchanged, i.e., $p_T(f|s)=p_R(f|s)$, for a gender-balanced 
population ($\beta=50\%$), as expected. 
If one considers where each curve crosses the line $p_T(f|s)=50\%$ (i.e., the gender classification of name $s$ is as good as a coin toss),
if follows that any name can be considered ambiguous in $\mathbf{T}$ if the group is skewed enough towards a gender (more specifically, if $\beta = 1-p_R(f|s)$). 
This property of the leaky pipeline transformation illustrates the challenge of classifying the gender of a name in situations of high gender bias. 
For example, the pink dotted line in Fig.~\ref{fig:cond_prob_functions}(a), corresponding to a female name with 99\% probability 
in $\mathbf{R}$, would provide a completely undefined gender guess if $\mathbf{T}$ comprises only 1\% females. 
Finally, the best gender guess for names equally shared between females and males in $\mathbf{R}$ (green dashed line), i.e., 
those for which $p_R(f|s)=50\%$, 
simply follows the gender profile of the group (i.e., $p_T(f|s)=\beta$). 
Hence in gGEM the gender classification of unisex names is completely steered by the context of the name list, not by the population at large.
Panel \emph{(B)} illustrates the transformation of Eq.~(\ref{eq:relation_condprob_1}) for four different gender mixes in $\mathbf{T}$. 
Once more, a gender-balanced population $\mathbf{T}$ (dashed green line), as a fair sample of $\mathbf{R}$, follows the same set of conditional probabilities. 
Changes are proportionately more dramatic as $\beta$ departs from $50\%$ either way. 
The particular case of $\beta=1\%$ (pink dotted line) illustrates that even names with strong gender association ($p_R(f|s)>99\%$)
should be considered unisex in the context of $\mathbf{T}$ if its gender profile is highly biased.

\subsection*{Self-consistency condition}

The conditional probabilities given by Eqs.~(\ref{eq:relation_condprob_1})--(\ref{eq:relation_condprob_2})
describe the gender profile of $\mathbf{T}$ if the social dynamic $\eta$ is known. 
However, this knowledge is not available \textit{a priori}. 

The gGEM framework posits a solution for $\eta$ that is consistent both with the 
list of names $\{N(s)\}$ observed in $\mathbf{T}$ and with the gender-name associations $\{p_R(g|s)\}$ known from $\mathbf{R}$
as if they were linked by the leaky pipeline.

A derivation of the self-consistent condition follows from the definition of any global parameter.
Choosing $\alpha$ for concreteness, we rewrite Eq.~(\ref{eq:def_alpha}) as $N_f-\alpha N_m = 0$. 
In addition, we use the identity $N_g=\sum_sp_T(g|s)N(s)$ to write:  
\begin{equation}
\sum_s \bigl( p_T(f|s) - \alpha \,p_T(m|s) \bigl) N(s) = 0\;.
\label{eq:selfconsistency_alpha}
\end{equation}
Self-consistency is imposed by noticing that the conditional probabilities that fulfill the leaky pipeline dynamic must follow Eqs.~(\ref{eq:relation_condprob_1})-(\ref{eq:relation_condprob_2}), which we substitute in Eq.~(\ref{eq:selfconsistency_alpha}). 
Moreover, the leaky pipeline equations [Eqs.~(\ref{eq:def_transf_biased_pop_1})-(\ref{eq:def_transf_biased_pop_2})] imply, by taking their ratio, that $\eta=\alpha$.
Elementary algebraic steps thus yield the condition, compactly written in terms of the gender imbalance $\gamma$ as:
\begin{equation}
\sum_{s} \frac{\delta_R(s)}{1+\delta_R (s)\,\gamma} N(s) = 0 \,,
\label{eq:cond_final}
\end{equation}
where $\delta_R(s)$, the \textit{gender-name inclination} in $\mathbf{R}$, is defined as
\begin{equation}
\delta_R (s) = p_R(f|s) - p_R(m|s) \;.
\label{eq:gender_definition}
\end{equation}
Eq.~(\ref{eq:cond_final}) is easily solved numerically for $\gamma$ by locating the sole zero-crossing value\footnote{There is no zero crossing for extreme pipelines producing $\mathbf{T}$ comprised only of females or males. In this case, the left-hand side of 
Eq.~(\ref{eq:cond_final}) is either strictly positive for female-only $\mathbf{T}$ (hence the solution is $\gamma=1$) or strictly negative for male-only $\mathbf{T}$ (hence $\gamma=-1$).} of  
its monotonically decreasing left-hand side\footnote{The self-consistent condition in $\gamma$ considering any value $\alpha^*\neq1$ reads as:
\begin{equation}
\sum_{s} \frac{\delta_R (s) -\gamma^*}{1-\gamma^* \delta_R (s) + [\delta_R (s) - \gamma^*]\gamma} N(s) = 0 \,,
\label{eq:cond_final_alphastar}
\end{equation}
where $\gamma^* = (\alpha^*-1)/(\alpha^*+1)$ is the gender imbalance of $\mathbf{R}$.
}. 

The gender profile solution provided by Eq.~(\ref{eq:cond_final}) uses all available names in the pool, including those with low gender-name inclination. 
When interpreted as a sum over broad classes of names sharing similar gender-name inclinations $\delta_R(s)$ (i.e. a sum over $\delta_R$ instead of over $s$), 
the expression can be understood as harnessing the fact that missing names in one gender tilt the balance of the gender distribution estimate towards the other. 
In this case, unisex names $(p_R(f|s) \approx p_R(m|s)\approx50\%$) do not contribute 
significantly to the estimate (since $\delta_R(s)\approx0$), as they carry little information about gender
(cf. green dashed curves in Fig.~\ref{fig:cond_prob_functions}\emph{(A)} and \emph{(B)}). 
Conversely, maximum information is provided by names that show high correlation with a given gender, 
i.e. those for which  $|\delta_R(s)|\approx1$. 
Names with intermediate vales of $\delta_R(s)$ contribute partial information (see {\em Materials and Methods} for derivations of the 
self-consistent condition based on other global parameters or on information theory.).

\section*{Performances of iGEMs and gGEM}

We tested the performance of gender estimation methods using lists of names artificially-generated in the computer
to build fictitious, or `synthetic', populations $\mathbf{T}$, each with well-controlled gender profile $\beta_0$. 

We employed three publicly available datasets in our simulations: The U.S. Social Security Administration's 2020 list of baby names~\cite{dataset_US}, 
Brazil 2010 census~\cite{dataset_BR}, and France 2019 INSEE's list of baby names~\cite{dataset_FR}. 
Each of these lists provided a reference population of the order of $10^8$ individuals from which the associated conditional probabilities $\{p_R(g|s)\}$ were extracted. 
We refer to them as $\mathbf{R_{[US]}}$, $\mathbf{R_{[BR]}}$, and $\mathbf{R_{[FR]}}$, respectively, 
and treat them as independent sets unless specified otherwise. 
To simplify notation, we refer to a generic reference population among them as $\mathbf{R_{[X]}}$, 
where $\mathbf{X}$ stands for $\mathbf{US}$, $\mathbf{BR}$, or $\mathbf{FR}$.

\begin{figure}[!tb]
\center{\includegraphics[width=0.98\columnwidth]{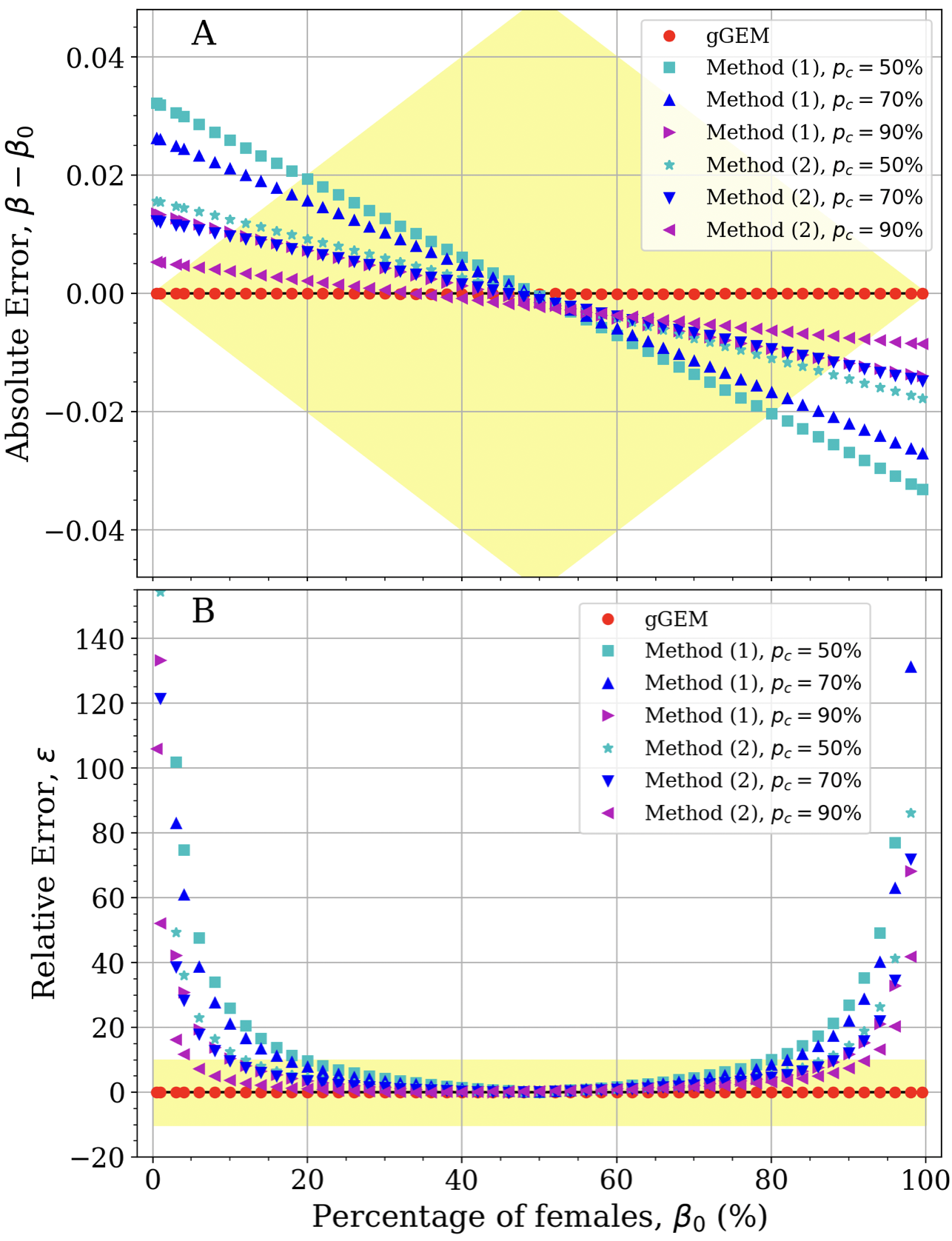}}
\caption{\label{fig:betas_US} Absolute (top) and relative (bottom) errors in the average estimate $\beta$ for the female fraction ($0\%<\beta<100\%$),
as functions of the `true' female fraction $\beta_0$ in the interval $0.5\%\leq\beta_0\leq99.5\%$, for different gender estimation methods. 
Synthetic populations $\mathbf{T_{[US]}}$ are used. 
Error bars denote the statistical uncertainty $\sigma_\beta$ over one thousand synthetic populations and have approximately the size of symbol markers in this case. 
The dashed black line marks the zero-error reference. The shaded yellow area denotes the 10\% relative confidence interval (with respect to the minority gender).
\emph{(A)}: absolute error, $\beta-\beta_0$. 
\emph{(B)}: relative error, $\varepsilon (\%) = 100\times[\min(\beta,1-\beta)-\min(\beta_0,1-\beta_0)] / \min(\beta_0,1-\beta_0)$.
} 
\end{figure}

Synthetic populations $\mathbf{T_{[X]}}$ 
were generated by sampling names from the same dataset used to generate the corresponding reference population $\mathbf{R_{[X]}}$ (see {\it Materials and Methods} for details).
We also considered large synthetic populations comprising ten thousand names each. 
This setting provides performance tests of the different methods in the most favorable conditions. 
The results of this section thus represent the best performance achievable, limited only by intrinsic methodological shortcomings of the inference strategies. 
Deviations from this ideal scenario are investigated in the next section. 

Each synthetic population was analyzed using Method~(1) [see Eq.~(\ref{eq:method_i})] and Method~(2) [see Eq.~(\ref{eq:method_ii})] with three values of cutoff probability commonly found in the literature\footnote{We note that Method (1) with $p_c=50\%$ coincides with Method (0).} ($p_c=50\%$, $p_c=70\%$, and $p_c=90\%$), and gGEM, 
yielding seven independent estimates $\beta$ for each dataset $\mathbf{[X]}$. For each method and each value of $\beta_0$, the average of 
estimates over a set of 1000 synthetic populations was adopted as the typical estimate $\beta$ provided by the method, while their standard deviation 
yielded the statistical uncertainty $\sigma_\beta$ denoted as error bars in the plots. 

Figure~\ref{fig:betas_US}\emph{(A)} depicts the error $\beta -\beta_0$ in the gender profile estimates of synthetic populations $\mathbf{T_{[US]}}$. 
All methods converge to the correct estimate for gender-balanced populations ($\beta_0\approx 50\%$) with error smaller than 1~percentage point (p.p.), as expected. 
However, estimates from Method~(1) and Method~(2) deviate linearly from the correct values as gender 
imbalance increases (i.e., as $\beta_0\rightarrow0\%$ or $\beta_0\rightarrow100\%$), uncovering a systematic methodological error. 
The choice of cutoff probability $p_c$ influences the linear sensitivity of the methods to this error source but cannot eliminate it completely. 
This is the quantitative consequence of the fair sampling hypothesis. 
Method~(2) with high cutoff probability fares better among iGEMs strategies, reaching better than 1~p.p. accuracy over the whole range of values $\beta_0$ in this ideal scenario.
gGEM does not suffer from intrinsic methodological issues, producing accurate estimates for all values of $\beta_0$.
The typical absolute error lies close to 0.01~p.p., i.e., compatible with finite size effects of the synthetic populations (ten thousand names). 
Similar trends are observed for $\mathbf{T_{[BR]}}$ and $\mathbf{T_{[FR]}}$. 

Relative errors are depicted in Fig.~\ref{fig:betas_US}\emph{(B)}. 
We consider errors relative to the fraction of the minority gender, since this is usually the quantity of interest in gender estimation. 
It can be either female for $\beta_0<50\%$, or male for $1-\beta_0<50\%$, 
and is denoted as $\min(\beta_0,1-\beta_0)$. 
The relative error is then defined as $\varepsilon (\%)= 100\times[\min(\beta,1-\beta)-\min(\beta_0,1-\beta_0)]/\min(\beta_0,1-\beta_0)$. 
Method~(1)'s zone of $10\%$ accuracy level (shaded yellow region), which we define as our standard confidence interval, is limited to 
around $20\%<\beta_0<80\%$ if all names are used ($p_c=50\%$). 
Method~(2) fares slightly better under the same conditions. 
By pushing $p_c$ up to 90\%, Method (2) is in principle able to achieve 10\% relative accuracy in the extended range $5\%<\beta_0<95\%$. 
In contrast, gGEM accuracy does not depend on $\beta_0$.

\begin{figure}[!tb]
\center{\includegraphics[width=0.99\columnwidth]{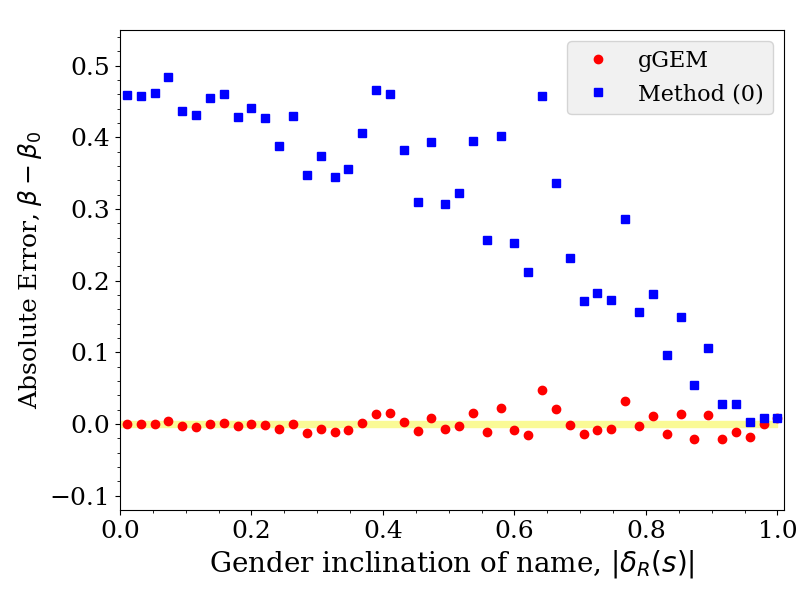}}
\caption{\label{fig:betas_Deltap} Partial contribution to the estimated female fraction in $\mathbf{T_{[BR]}}$ stemming from each group of names with given 
gender-name inclination $|\delta_R(s)|$ (absolute value),  according to gGEM [red circles] and Method (0) [blue squares]. The $\beta$ estimates for the whole population 
are indicated by the dashed blue line for gGEM and dotted red line for Method (0). The input synthetic population is composed of 4\% females ($\beta_0 = 0.040$). 
The shaded yellow area indicates the 10\% relative confidence interval.} 
\end{figure}

Figure~\ref{fig:betas_Deltap} illustrates how gGEM (red circles) and iGEM (blue squares) treat unisex and gender-defined names differently to build an estimate. 
It shows the partial contributions of subsets of names grouped according to gender-name inclination $|\delta_R(s)|$. 
A single synthetic population $\mathbf{T_{[BR]}}$ composed of $10^5$ names and 4\% female ($\beta_0=4\%$) was analyzed.
gGEM produces the estimate $\beta_{gGEM}=4.1\%$, while Method~(0) overestimates females presence by almost 50\%, at $\beta_{iGEM}=5.7\%$. 
The discrepancy between methods is caused by the different mechanisms of gender allocation. 
Method~(0) (blue squares) works by distributing every subgroup in $\mathbf{T_{[BR]}}$ following the gender-name inclinations observed 
in $\mathbf{R_{[BR]}}$ [see Eq.~(\ref{eq:method_i})]. 
In particular, subpopulations bearing names with undefined gender ($\delta_R(s)\approx0$) are equally split between the two genders. 
This effect, which becomes more pronounced for unisex names, is what causes overestimation of the minority gender participation in iGEM methods. 
The size of the systematic error is hence proportional to the fraction of people bearing unisex names. 
For Western populations, most names have well defined gender, but that is not the case in several other regions of the world~\cite{huang2020historical}.
Methods~(1) and~(2) improve the situation by simply disregarding the contributions from people bearing names for which $|\delta_R(s)|\leq p_c$.  
By contrast, the gGEM self-consistency condition of Eq.~(\ref{eq:cond_final}) imposes that all subsets of names, regardless of 
gender-name inclination, produce the same $\beta$ estimate globally. 
In fact, the useful information that each group of names carries about gender is proportional to $|\delta_R(s)|$, 
which can be understood as the `sensitivity' of the name to the leaky pipeline dynamic. 
gGEM thus weighs each subset of names according to its sensitivity to gender and proportionally disregards ambiguous contributions 
that would introduce systematic errors in favor of what the global ensemble of names indicates.

\section*{Robustness of Gender Estimates}

\subsection*{Mismatched reference datasets}

We now investigate the robustness of gender estimation methods with respect to the choice of reference population. 
Making an optimal choice is indeed nontrivial because target populations are shaped by broad social factors, making 
first names not only correlated with gender, but also with other identity traits related to the origin of individuals, 
such as geographic region or country of birth, religious community, or even socio-economic background. 

\begin{figure}[!tb]
\center{\includegraphics[width=0.99\columnwidth]{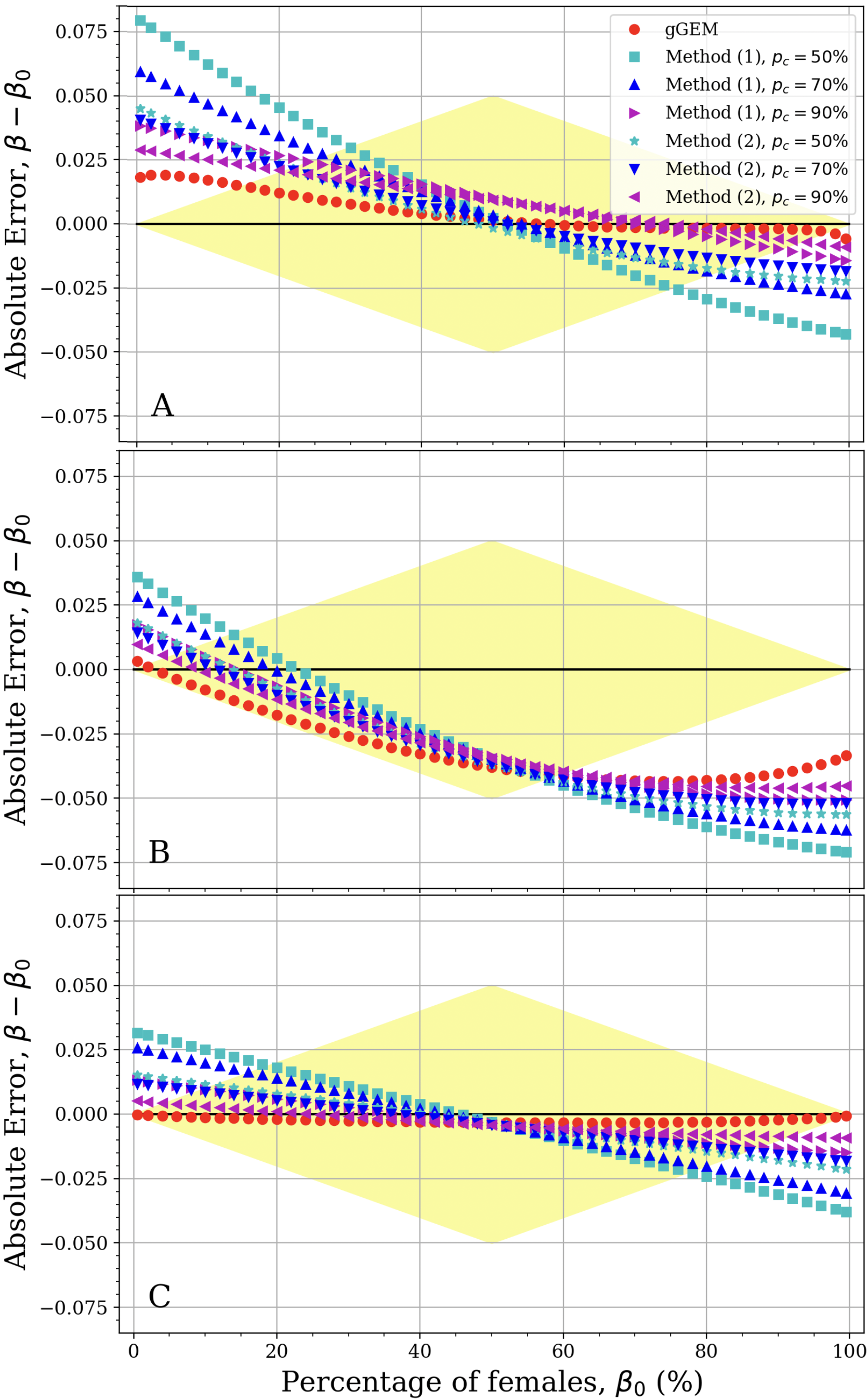}}
\caption{\label{fig:robutsness_betas_synthpopUS} Error in gender estimate $\beta$ with respect to gender composition $\beta_0$ of target population for mismatched reference population. 
{\em (A)} Synthetic populations $\mathbf{T_{[FR]}}$ analyzed with reference population $\mathbf{R_{[BR]}}$. 
{\em (B)} $\mathbf{T_{[US]}}$ analyzed with $\mathbf{R_{[BR]}}$. 
{\em (C)} $\mathbf{T_{[US]}}$ analyzed with the union $\mathbf{R_{[all]}}$ of three reference populations. 
Yellow-shaded areas delimit the region where gender estimates are within 10\% relative error of the minority gender. Dashed black line marks the zero-error reference.} 
\end{figure}

In this section, we analyze synthetic target populations $\mathbf{T_{[X]}}$ using a mismatched reference population $\mathbf{R_{[X']}}$ (with $X'\neq X$). 
We expect with this to reproduce more realistic scenarios in which there are systematic differences 
between $\mathbf{T_{[X]}}$ and $\mathbf{R_{[X']}}$ in the frequency of names and their gender-name inclination,
so as to provide a quantitative characterization of the role played by these effects and ways to mitigate them. 

Figure~\ref{fig:robutsness_betas_synthpopUS}\emph{(A)} illustrates the analysis of French 
synthetic populations $\mathbf{T_{[FR]}}$ using as reference the Brazilian population in $\mathbf{R_{[BR]}}$. 
As a first limitation, only 29\% of all unique names are detected on average (i.e. about 71\% of unique names do no appear in $\mathbf{R_{[BR]}}$). 
Since these tend to be common names in both populations, the number of individuals detected is still high, at approximately 74\% for gGEM (and iGEMs with $p_c=50\%$),
and 66\% for iGEMs with $p_c=90\%$. 
Furthermore, identified people are fairly distributed between genders, putting to rest one possible mechanism of estimation bias in this particular case. 
The figure shows that all methods are prone to systematic errors that increase as $\mathbf{T_{[FR]}}$ departs from gender-balance. 
gGEM retrieves the most accurate estimates, reaching uncertainty at the 2~p.p. level. 

Since several names are shared between Brazilian and French populations due to common linguistic roots, 
one can be expected to still provide gender information of good quality about the other. 
Figure~\ref{fig:robutsness_betas_synthpopUS}\emph{(B)} presents a situation in which target and reference populations of names have less in common: US populations $\mathbf{T_{[US]}}$ are analyzed using the 
Brazilian reference population $\mathbf{R_{[BR]}}$. 
Only 16\% of all unique names are now identified, although still corresponding to a large fraction of all individuals on average (73\% for gGEM and $68\%$ for iGEMs with $p_c=90\%$). 
However, they are now unevenly distributed between genders, as less women (70\%) are matched than men (76\%) for gGEM, introducing a source of bias in the estimation.
In fact, errors now reach about 5~p.p. for all methods and depart from linearity in $\beta_0$ to show clear asymmetry between male- and female-dominated target populations.

These results suggest to use the fractions of recognized people and names 
as possible proxies for gauging the appropriateness of the reference dataset given a target population. 
The figures of about at least 75\% of individuals identified and 30\% of unique names matched seem to 
provide in our tests a lower bound for the reliability of gender estimation.
Conversely, a high rate of identification for both names and individuals is good indication that high accuracy is achievable.
We observed results compatible with the best methodological accuracy of last section when at least 90\% of individuals are identified.

A possible solution to mitigate population mismatch issues consists in oversampling the reference population. 
In a final analysis, we combined the three datasets, $\mathbf{[US]}$, $\mathbf{[BR]}$ and $\mathbf{[FR]}$, into a 
single reference population $\mathbf{R_{[all]}}$\footnote{Depending on how much information is available about the probable origin of the target population, other strategies could be adopted to combine reference datasets. 
Our strategy of simply combining all individuals in a single pool to recalculate conditional probabilities aims to preserve the gender-name correlations of names that are frequent in each dataset.}.
This ensures 100\% name recognition (and of individuals) for gGEM (or 94\% of individuals for iGEMs with $p_c=90\%$).
The resulting analysis, shown in Fig.~\ref{fig:robutsness_betas_synthpopUS}\emph{(C)}, indicates that gender estimation 
of $\mathbf{T_{[US]}}$ using the oversampled $\mathbf{R_{[all]}}$ provides much more reliable estimates overall 
than those obtained in panel \emph{(B)}. 
They are also in fair agreement with those obtained in Fig.~\ref{fig:betas_US}, where the ideally matching reference 
population $\mathbf{R_{[US]}}$ was used. 
gGEM's uncertainty stands at better than the 0.5 p.p. level in this case.

The improvement produced by oversampling $\mathbf{R}$ occurs because subsets of the most frequent names 
for $\mathbf{[US]}$, $\mathbf{[BR]}$ and $\mathbf{[FR]}$ possess either small overlap or little disagreement in gender classification, 
a consequence of the \textit{sparseness} of the sampling in $s$. 
In this situation, conditional probabilities in $\mathbf{R_{[all]}}$ are dominated by the reference set in which they represent the most frequent names for datasets of similar sizes. 

\begin{figure}[!tb]
\center{\includegraphics[width=0.99\columnwidth]{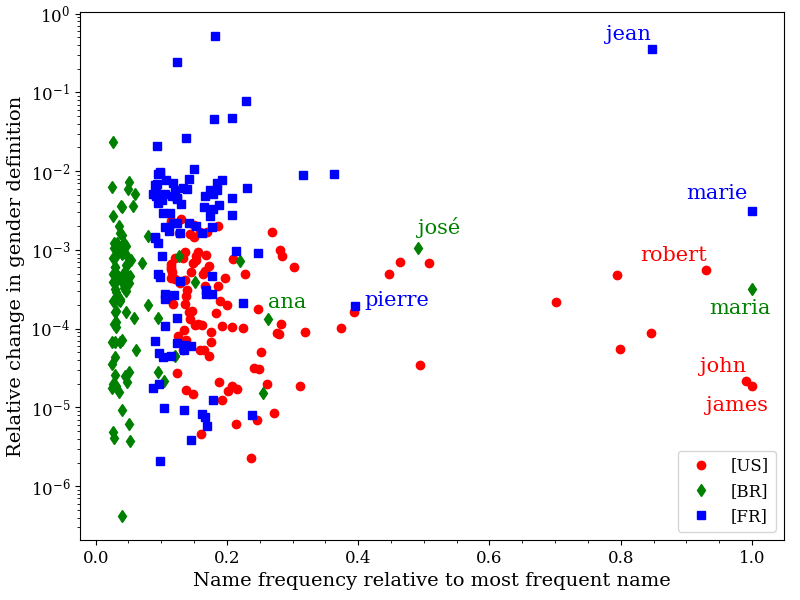}}
\caption{\label{fig:robutsness_condprobs} Relative change in gender-name inclination  $\sigma_\mathbf{R_{[X]}}(s)$ for the top-100 names in each country-wide reference population.
The three most frequent names of each dataset are indicated. } 
\end{figure}

Figure~\ref{fig:robutsness_condprobs} supports this intuition. 
It depicts how conditional probabilities for the top-100 names in each country-wide dataset change in $\mathbf{R_{[all]}}$ with respect 
to values in their original reference population $\mathbf{R_{[X]}}$. 
The plot shows absolute values of the relative change in gender-name inclinations, defined for each name as $\sigma_\mathbf{R_{[X]}}(s)=|\delta_\mathbf{R_{[all]}}(s)-\delta_\mathbf{R_{[X]}}(s)|/|\delta_\mathbf{R_{[X]}}(s)|$, 
portrayed as functions of the name frequency in $\mathbf{R_{[X]}}$ (relative to the most frequent name). 
The plot indicates that the most frequent names undergo little change in gender-name inclination by oversampling country-wide human populations, 
with typical relative differences fulfilling $\sigma_\mathbf{R_{[X]}}(s) < 1$\%. 
Name `collision' effects that could substantially affect gender-name conditional probabilities are thus rare~\footnote{One remarkable exception is the second most common name in 
France, $s_2=$``Jean'', a clearly male name in that country ($\delta_\mathbf{R_{[FR]}}(s_2) < -0.999$). 
Even though it does not rank among the top-100 in the US (where it has little impact in 
gender estimates), its frequency and discrepancy in gender-name inclination ($\delta_\mathbf{R_{[US]}}(s_2) =0.883$) in this country are high enough to alter its gender-name inclination 
in $\mathbf{R_{[all]}}$. Moreover, the US dataset is larger. }.

The sparseness condition that allows us to combine different datasets without `collisions' can also be stated in terms of the observed relatively low diversity of names in a typical reference population. 
The Shannon entropy of $s$ in $\mathbf{R_{[X]}}$, calculated from the name frequencies $p(s)$ as $H=-\sum_s p(s)\log p(s)$, is $\approx10$ bits. 
This roughly means that a subset of about only one thousand frequent names ($\approx2^{10}$) in $\mathbf{R_{[X]}}$ accounts for most of the information carried by the symbol $s$ in a population of $\approx10^8$ individuals. 

Moreover, this property of $\mathbf{R_{[X]}}$ also explains why precise probabilities $p(s)$ of name occurrence are not consequential to gender estimation (in fact, they are never a concern in the literature).
For target populations $\mathbf{T}$ that are not several orders of magnitude larger than the typical set of names ($\approx2^{10}$) in $\mathbf{R}$, first names assume the characteristics of a random variable. 
This means that little information about the specific values $p(s)$ remains in $\mathbf{T}$, thus justifying the fact that precise knowledge of the set $\{p(s)\}$ is not required for reliable gender estimation. 
Most names from $\mathbf{R}$ will indeed not be present in $\mathbf{T}$, and those that are present will typically appear only a few times.

\subsection*{Datasets with limited gender information}

So far, we have used first names as a collection of symbols 
that, via correlations with gender, provide information about the gender composition of a population.
We now investigate situations wherein the information at hand is severely limited:
only the first or the last letter of each name is available.

\begin{figure}[!tb]
\center{\includegraphics[width=0.99\columnwidth]{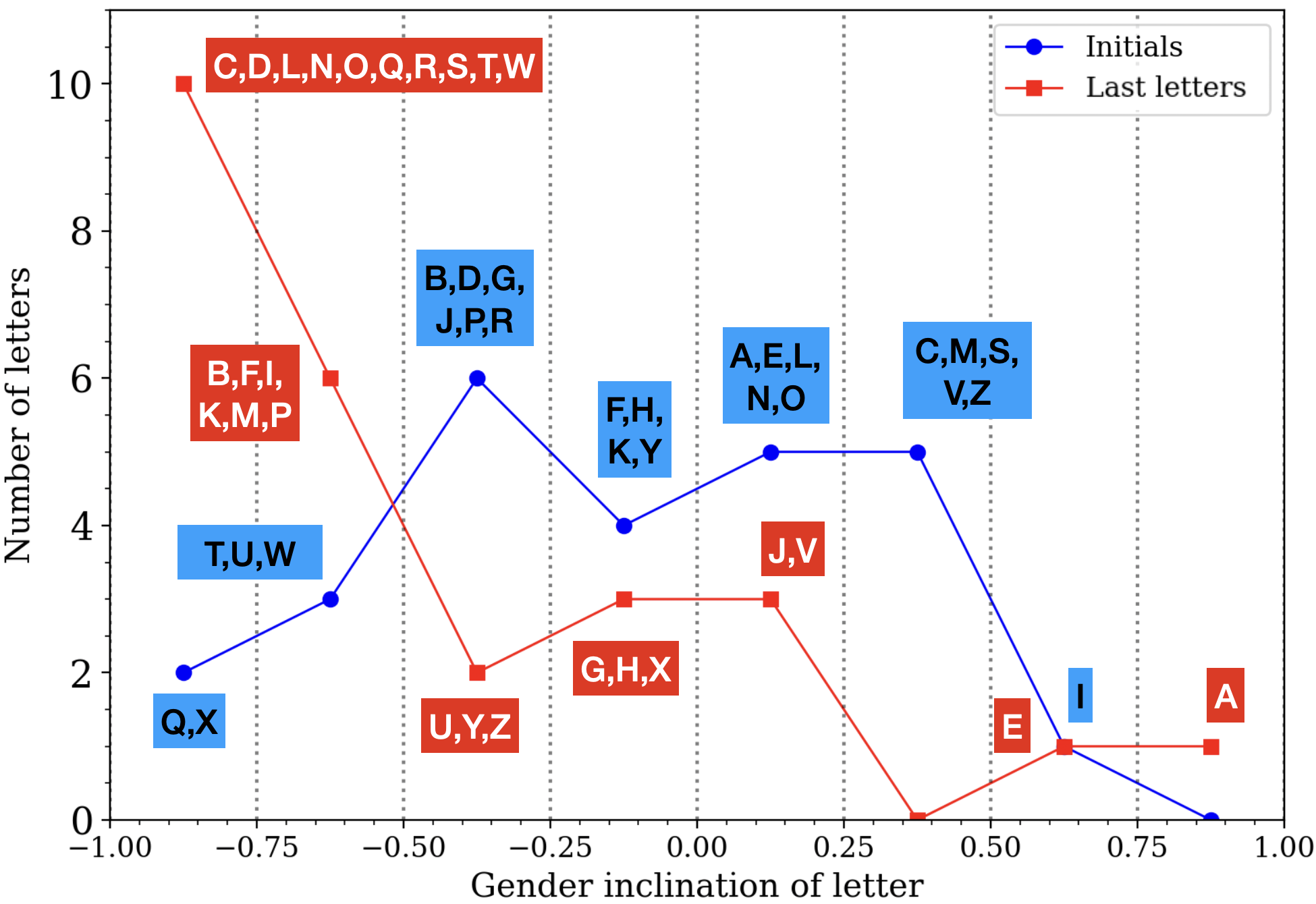}}
\caption{\label{fig:hist_letras_IF_FR} Frequency of letters in the Latin alphabet for a given gender-name inclination in $\mathbf{R_{[FR]}}$ considering 
only initials ($\delta_R^I(s)$, blue circles) and last letters ($\delta_R^L(s)$, red squares). The dashed vertical lines delimit the bins we used in the plot and markers 
are placed in the middle of each bin. 
} 
\end{figure}

Statistics on how initials and last letters correlate with gender are retrieved from a reference population $\mathbf{R}$ in the same way one does for first names.
The symbol $s$ in this case denotes one of the 26 letters in the Latin alphabet once diacritics are converted to the nearest form (e.g., \c{c}$\,\rightarrow\,$c). 
Two new sets of conditional probabilities follow: $\{p_R^I(g|s)\}$ for first-name initials and $\{p_R^L(g|s)\}$ for end letters. 
Gender inclinations of letters, calculated as before, are now respectively denoted as $\delta_R^I(s)$ and as $\delta_R^L(s)$.

\begin{figure}[!b]
\center{\includegraphics[width=0.99\columnwidth]{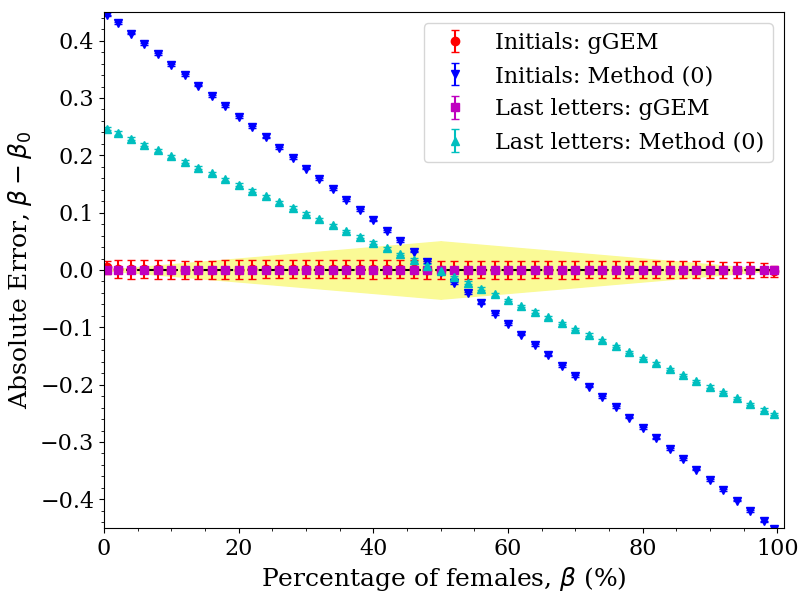}}
\caption{\label{fig:betas_IF_FR} Error of female fraction estimates, $\beta-\beta_0$ using 
initial or last letters of synthetic populations generated from $\mathbf{R_{[FR]}}$,
for both iGEM Method~(0) and gGEM. 
The shaded yellow region indicates the 10\% relative confidence interval.} 
\end{figure}

Histograms of gender inclination of initial and last letters for the French reference set $\mathbf{R_{[FR]}}$
are shown in Fig.~\ref{fig:hist_letras_IF_FR} (the other reference populations, $\mathbf{R_{[US]}}$ and $\mathbf{R_{[BR]}}$, follow similar patterns). 
Initials present a broad peak around zero gender-letter inclination ($\delta_R^I(s)=0$), indicating a low but nonzero correlation with gender. 
In contrast, the gender-letter inclination $\delta_R^L(s)$ of last letters decreases monotonically, indicating that several letters 
(in fact, about half of the alphabet) are strongly indicative of male gender while very few are exclusive to female gender. 
While first names are typically highly correlated with gender, initial and last letters provide very limited gender information if taken in isolation. 
iGEMs are thus expected to perform poorly in this context. 
We compared the performance of letter-based Method~(0) and gGEM. 
(The cut-off condition renders Methods~(1) and ~(2) useless for initials, 
or inapplicable for last letters, as empty sets or male-only sets are obtained as reference population, so we do not consider them here). 

Results of the analysis of synthetic populations following the same 
procedures as before are shown in Fig.~\ref{fig:betas_IF_FR}. 
As expected, Method~(0) is indeed nearly insensitive to the gender composition of the population
for initials (blue down-triangles) and weakly sensitive for last letters (cyan up-triangles):
target populations are always estimated to be close to gender-balance regardless of their gender profile $\beta_0$. 

gGEM attains once more reliable gender estimates, albeit with much larger statistical uncertainty than those obtained using first names.
The method remains accurate (no systematic errors), but statistical error bars (1 standard deviation) are now visible at the 0.5~p.p. level for 
last letters (magenta squares) and 1.5~p.p. level for initials (red circles),
rendering gGEM less precise in the ideal scenario. 
This is due to the smaller amount of gender information provided by single letters. 
End letters are also better distributed between genders, yielding information of better quality. 
These examples show that gGEM is able to produce reliable estimates even if only weak correlations are available between 
the property being estimated (gender profile) and the symbols to which it is associated (initials or final letters). 
We note, however, that matching target and reference populations becomes more important in this case, as the decreased amount of 
redundancy in gender-letter correlations makes the method less robust to mismatch.

\section*{Discussion and Conclusion}

The estimation of the gender profile of a group from names involves two elements: 
(i) the diligent choice of a reference dataset, from which gender-name
associations can be obtained, and (ii) a procedure that connects the list of names to this source of statistics. 

By dealing with element (i), current `individual-based gender estimation methods' (iGEMs) focus on improving the quality of 
the reference dataset and the sophistication with which information can be harnessed to `genderize' each name. 
While an important effort, this strategy in isolation can only achieve limited accuracy and is insufficient to improve the 
quality of gender profile estimation significantly. 

In this paper, we tackled aspect (ii), i.e., the method by which individual gender labels are interpreted 
to produce a gender estimate for a target group. 
Based on a leaky pipeline model of social dynamic, 
the `global gender estimation method' (gGEM) introduced here is logically consistent and takes into account the context of the name list of interest.

A performance study under ideal conditions established that gGEM is free of intrinsic methodological systematic effects, 
an important conceptual property of any reliable measurement tool. 
iGEMs, on the other hand, systematically underestimate gender imbalance, 
particularly for groups with strong underrepresentation of the minority gender, i.e., the case of most interest.
We provided lower bounds for the region of parameters within which iGEMs can be considered unreliable due to intrinsic methodological shortcomings, revealing
that they are expected to surely lose accuracy when the minority gender comprises less than about 10\% of the population. 

When facing more realistic scenarios, gender estimation is also much more robust with gGEM than with iGEMs. 
Practical limitations such as misidentification of gender labels and lack of gender information for certain names were investigated by
using a reference population that is knowingly inappropriate (mismatched) to analyze the gender profile of synthetic populations. 
Oversampling the reference population is suggested as a practical mitigation measure against these issues. 
For large populations, our simulations indicate that gGEM should be accurate to better than 1~p.p. level for any mix of genders, 
while iGEMs show degraded accuracy if the minority gender comprises about 20\% of the group or less. 

One of the methodological strengths of gGEM lies on the optimal use of partial gender-name information. 
This makes the method suitable to analyze populations with a high share of unisex names, 
such as those in East Asia, without compromising accuracy or breath of name recognition. 
As a consequence, gGEM also performs well in more general situations wherein little information 
(e.g. first-name initials) is available. 
This feature has the potential to extend the usage of gGEM variants to other types of inference based on weakly correlated data. 

Finally, and importantly in practice, gGEM is simple to implement, 
and does not require any significant computing power: 
One merely needs to input the name-gender probabilities of the reference population and the name counts of the target group into Eq.~(\ref{eq:cond_final}), 
and vary $\gamma$ to find the value at which Eq.~(\ref{eq:cond_final}) is satisfied, 
which is nothing but the best estimate of the departure from gender-parity of the group of interest.

Given its fundamental superiority and implementation simplicity, we expect gGEM and
gGEM-like approaches to be widely used for gender profile estimation and other problems where collective properties
need to be estimated from partial information on individual members of the ensemble.

\matmethods{

\subsection*{Self-consistency conditions}

Other equivalent conditions also lead to Eq.~(\ref{eq:cond_final}). We outline key steps in alternative derivations, assuming $\alpha^*=1$ for simplicity. 
A self-consistent consistent condition in $\beta$ equivalent to Eq.~(\ref{eq:selfconsistency_alpha}) stems from the identity $\beta (N_f+N_m)-N_f=0$. 
A self-consistent condition in $\gamma$ uses the identity $N_f - N_m - \gamma (N_f+N_m)=0$. In both cases, replacing $N_g$ by the identity $N_g=\sum_sp_T(g|s)N(s)$ 
and using the transformation of Eq.~(\ref{eq:relation_condprob_1}) leads to Eq.~(\ref{eq:cond_final}). 
Alternatively, one may impose that the error $\epsilon(s)= p_T(f|s) N(s) - N_f(s)$ in the estimate of the number of females bearing name $s$ be zero on average in $s$, 
through the condition $\sum_{s}\epsilon(s) = 0$. Also this  leads to Eq.~(\ref{eq:cond_final}). A different self-consistent condition may be imposed on the Shannon 
entropy of gender $H_g$, which on one side is a function of a global gender parameter (say, $\beta$) through $H_g(\beta) = -\beta \log\beta - (1-\beta) \log(1-\beta)$ and, 
on the other, must transform according to Eqs.~(\ref{eq:relation_condprob_1})-(\ref{eq:relation_condprob_2}) as $H_g(\alpha) = -\sum_{g,s}p_T(g|s)\log p_T(g|s)$. 
Imposing $H_g(\beta) = H_g(\alpha)$ produces once more the same self-consistent condition of Eq.~(\ref{eq:cond_final}).

\subsection*{Generation of synthetic populations}

We generate fictitious or synthetic populations with well-controlled gender profile $\beta_0$ in the following fashion.
First, each dataset $\mathbf{[X]}$ was separated into two subsets, a female-only and a male-only. 
Second, a target population $\mathbf{T_{[X]}}$ with the well-defined gender profile $\beta_0=N_f/(N_f+N_m)$ was generated by randomly 
sampling the specified number $N_f$ of names from the female-only subset and equivalently for $N_m$ names from the male-only subset such that 
a total of $N_f+N_m=10000$ names were sampled. 
Sampling was performed respecting the natural frequency of names (fair sampling). 
We tested whether sampling every unique name with the same probability (uniform sampling) would affect estimation performance and concluded that quantitative 
differences result from the fact that in this setting unisex names assume higher weight. 
To characterize fluctuations, the second step described above was repeated one thousand times to generate 1000 independent 
populations $\mathbf{T_{[X]}}$ for the same value of $\beta_0$.
Finally, 52 different values of $\beta_0$ representing populations with female participation ranging from 0.5\% to 99.5\% were chosen 
to generate the plots that follow.
Reference populations $\mathbf{R_{[X]}}$ exclude names with less than 100 individuals, since conditional probabilities
$p_R(g|s)$ in those cases have large uncertainty. 
Including such low-probability names increases statistical noise while bringing little increase in the number of individuals identified. 
}

\showmatmethods{} 




\begin{thebibliography}{10}

    \bibitem{helmer2017research}
    Helmer M, Schottdorf M, Neef A, Battaglia D (2017) Research: Gender bias in
      scholarly peer review.
    \newblock {\em eLife} 6:e21718.
    
    \bibitem{king2017men}
    King MM, Bergstrom CT, Correll SJ, Jacquet J, West JD (2017) Men set their own
      cites high: Gender and self-citation across fields and over time.
    \newblock {\em Socius} 3:2378023117738903.
    
    \bibitem{murray2019author}
    Murray D, et~al. (2019) Author-reviewer homophily in peer review.
    \newblock {\em bioRxiv}.
    
    \bibitem{dworkin2020extent}
    Dworkin JD, et~al. (2020) The extent and drivers of gender imbalance in
      neuroscience reference lists.
    \newblock {\em Nature Neuroscience} 23(8):918--926.
    
    \bibitem{chatterjee2021gender}
    Chatterjee P, Werner RM (2021) {Gender Disparity in Citations in High-Impact
      Journal Articles}.
    \newblock {\em JAMA Network Open} 4(7):e2114509--e2114509.
    
    \bibitem{squazzoni2021peer}
    Squazzoni F, et~al. (2021) Peer review and gender bias: A study on 145
      scholarly journals.
    \newblock {\em Science Advances} 7(2):eabd0299.
    
    \bibitem{teich2022citation}
    Teich EG, et~al. (2022) Citation inequity and gendered citation practices in
      contemporary physics.
    \newblock {\em Nature Physics} 18(10):1161--1170.
    
    \bibitem{huang2020historical}
    Huang J, Gates AJ, Sinatra R, Barab{\'a}si AL (2020) Historical comparison of
      gender inequality in scientific careers across countries and disciplines.
    \newblock {\em Proceedings of the National Academy of Sciences}
      117(9):4609--4616.
    
    \bibitem{ross2022women}
    Ross MB, et~al. (2022) Women are credited less in science than men.
    \newblock {\em Nature} 608(7921):135--145.
    
    \bibitem{dataset_US}
    (2022) https://www.ssa.gov/oact/babynames/limits.html.
    
    \bibitem{lariviere2013bibliometrics}
    Larivi{\`e}re V, Ni C, Gingras Y, Cronin B, Sugimoto CR (2013) Bibliometrics:
      Global gender disparities in science.
    \newblock {\em Nature} 504(7479):211--213.
    
    \bibitem{wais2016gender}
    Wais K (2016) {Gender Prediction Methods Based on First Names with genderizeR}.
    \newblock {\em {The R Journal}} 8(1):17--37.
    
    \bibitem{karimi2016inferring}
    Karimi F, Wagner C, Lemmerich F, Jadidi M, Strohmaier M (2016) Inferring gender
      from names on the web: A comparative evaluation of gender detection methods
      in {\em Proceedings of the 25th International Conference Companion on World
      Wide Web}, WWW '16 Companion.
    \newblock (International World Wide Web Conferences Steering Committee,
      Republic and Canton of Geneva, CHE), pp. 53--54.
    
    \bibitem{santamaria2018comparison}
    Santamaria L, Mijalhevic H (2018) Comparison and benchmark of name-to-gender
      inference services.
    \newblock {\em PeerJ Computer Science} 4:e156.
    
    \bibitem{fortin2021digital}
    Fortin J, Bartlett B, Kantar M, Tseng M, Mehrabi Z (2021) Digital technology
      helps remove gender bias in academia.
    \newblock {\em Scientometrics} 126(5):4073--4081.
    
    \bibitem{das2021context}
    Das S, Paik JH (2021) Context-sensitive gender inference of named entities in
      text.
    \newblock {\em Information Processing and Management} 58(1):102423.
    
    \bibitem{hu2021what}
    Hu Y, et~al. (2021) What's in a name? --gender classification of names with
      character based machine learning models.
    \newblock {\em Data Mining and Knowledge Discovery} 35(4):1537--1563.
    
    \bibitem{smith2013search}
    Smith BN, Singh M, Torvik VI (2013) A search engine approach to estimating
      temporal changes in gender orientation of first names in {\em Proceedings of
      the 13th ACM/IEEE-CS Joint Conference on Digital Libraries}, JCDL '13.
    \newblock (Association for Computing Machinery, New York, NY, USA), pp.
      199--208.
    
    \bibitem{ethnea2016torvik}
    Torvik V, Agarwal S (2016) Ethnea -- an instance-based ethnicity classifier
      based on geo-coded author names in a large-scale bibliographic database.
    \newblock International Symposium on Science of Science ; Conference date:
      22-03-2016 Through 23-03-2016.
    
    \bibitem{muller2017improving}
    M{\"u}ller D, Te YF, Jain P (2017) Improving data quality through high
      precision gender categorization in {\em 2017 IEEE International Conference on
      Big Data (Big Data)}.
    \newblock pp. 2628--2636.
    
    \bibitem{vanbuskirk2022open}
    Van~Buskirk I, Clauset A, Larremore DB (2022) An open-source cultural consensus
      approach to name-based gender classification.
    
    \bibitem{ggemapp}
    (2023) https://ggem.app
    
    \bibitem{thomas2019gender}
    Thomas EG, et~al. (2019) {Gender Disparities in Invited Commentary Authorship
      in 2459 Medical Journals}.
    \newblock {\em JAMA Network Open} 2(10):e1913682--e1913682.
    
    \bibitem{mattauch2020bibliometric}
    Mattauch S, Lohmann K, Hannig F, Lohmann D, Teich J (2020) A bibliometric
      approach for detecting the gender gap in computer science.
    \newblock {\em Commun. ACM} 63(5):74--80.
    
    \bibitem{dew2021gendered}
    Dew M, Perry J, Ford L, Bassichis W, Erukhimova T (2021) Gendered performance
      differences in introductory physics: A study from a large land-grant
      university.
    \newblock {\em Phys. Rev. Phys. Educ. Res.} 17(1):010106.
    
    \bibitem{dataset_BR}
    (2022) https://brasil.io/dataset/genero-nomes/files.
    
    \bibitem{dataset_FR}
    (2022) https://www.insee.fr/fr/statistiques/2540004?sommaire=4767262.
    
    \end{thebibliography}
\end{document}